\documentclass[a4paper,hyperref,option,numerical,linenumbers,superscriptaddress,nofootinbib,showpacs,showkeys]{PoS}

\usepackage{graphicx}
\usepackage{cite}
\usepackage{graphicx,color}
\usepackage{amsmath,amssymb,amsfonts}

\usepackage{lineno}

\def \be {\begin{equation}}
\def \ee {\end{equation}}
\def \ee  {\end{equation}}
\def \bea {\begin{eqnarray}}
\def \eea {\end{eqnarray}}

\newcommand{\pT} {\ensuremath{p_{\mathrm{T}}}}

\title{Beam energy dependence of the anisotropic flow coefficients v$_n$}

\ShortTitle{Beam energy dependence of the anisotropic flow coefficients $\mathrm{v_n}$}

\author{\speaker{Niseem Magdy} (For the STAR Collaboration)\\
        Department of Chemistry, Stony Brook University, Stony Brook, NY, 11794-3400, USA\\
        E-mail: \email{niseemm@gmail.com}}
        
\abstract{
Recent STAR measurements of the anisotropic flow coefficients, v$_{n}$, are presented for Au+Au collisions spanning the beam energy range $\sqrt{s_{NN}} = 7.7 - 200$~GeV. The measurements indicate dependences on harmonic number, $n$, transverse momentum ($p_T$), pseudorapidity ($\eta$), collision centrality ($\mathrm{cent}$) and beam energy ($\sqrt{s_{NN}}$) which could serve as important constraints to test different initial-state models and to aid precision extraction of the temperature dependence of the specific shear viscosity.
}

\FullConference{Critical Point and Onset of Deconfinement\\,
         7-11 August  2017\\
         The Wang Center, Stony Brook University, Stony Brook, NY} 

\usepackage{amsmath}

\begin{document}

\section{Introduction}

A major goal of the heavy-ion experimental program at the Relativistic Heavy Ion Collider (RHIC) is to study the properties of the strongly interacting quark-gluon plasma (QGP) created in ion-ion collisions. Recently, many studies have emphasized the use of anisotropic flow measurements to study the transport properties of the QGP ~\cite{Teaney:2003kp,Lacey:2006pn,Schenke:2011tv, Song:2011qa,Niemi:2012ry,Qin:2010pf,Magdy:2017kji}.  
An important question in many of these studies has been the role of initial-state fluctuations and their influence on the uncertainties associated with the extraction of $\eta/s$ for the QGP~\cite{Alver:2010gr,Lacey:2013eia}. This work presents new measurements for the anisotropic flow coefficients, v$_{n > 1}$, and the rapidity-even dipolar flow coefficient, v$^{even}_{1}$, with an eye toward developing new constraints which could aid a distinction between different initial-state models and hence, facilitate a more precise extraction of the specific shear viscosity $\eta/s$~\cite{Auvinen:2017fjw,Auvinen:2017pny}. 

Anisotropic flow is characterized by the Fourier coefficients,  v$_{n}$, obtained from a Fourier expansion 
of the azimuthal angle ($\phi$) distribution of the particles emitted in the collisions~\cite{Poskanzer:1998yz}:

\begin{eqnarray}
\label{eq:1}
\frac{dN}{d\phi}\propto1+2\sum_{n=1}\mathrm{v_{n}}\cos (n(\phi-\Psi_{n})),
\end{eqnarray}
where $\Psi_n$ represents the azimuthal angle of the $n^{th}$-order event plane; the coefficients, v$_{1}$, v$_{2}$ and v$_{3}$ are commonly called directed,  elliptic, and  triangular flow, respectively. The flow coefficients,  v$_{n}$, are related to the two-particle Fourier coefficients, v$_{n,n}$, as:

\begin{eqnarray}
\label{eq:3}
\mathrm{v_{n,n}}(\pT^{a},\pT^{b})  = \mathrm{v_n}(\pT^{a})\mathrm{v_n}(\pT^{b})+ \mathrm{\delta_{NF}},
\end{eqnarray}
where a and b are particles selected with $\pT^{a}$ and $\pT^{b}$ respectively, and $\mathrm{\delta_{NF}}$ is a so-called  non-flow (NF) term, which  includes possible short-range contributions from resonance decays, Bose-Einstein correlations and near-side jets, and long-range contributions from the global momentum conservation (GMC) ~\cite{Lacey:2005qq,Borghini:2000cm,ATLAS:2012at}. 
The short-range contributions can be reduced by employing a pseudorapidity gap, $\Delta\eta$. However, the effects of GMC must be explicitly considered. For the current analysis, a simultaneous fitting procedure, outlined below, was used to account for GMC.

\section{Measurements}
\label{Measurements}
The correlation function technique was used to measure the two-particle $\Delta\phi$ correlations:
\begin{eqnarray}\label{corr_func}
 C_{r}(\Delta\phi, \Delta\eta) = \frac{(dN/d\Delta\phi)_{same}}{(dN/d\Delta\phi)_{mixed}},
\end{eqnarray} 
where  $(dN/d\Delta\phi)_{same}$ represent the normalized azimuthal distribution of  particle pairs  from the same 
event and $(dN/d\Delta\phi)_{mixed}$ represents the normalized azimuthal distribution for particle pairs in which each member  is selected  from a different  event but with a similar classification for the collision vertex location,  centrality, etc. The pseudorapidity requirement  $|\Delta\eta| > 0.7$ was also imposed on track pairs to minimize non-flow contributions associated with the short-range correlations.

The two-particle Fourier coefficients, v$_{n,n}$, are obtained from the correlation function as:
\begin{eqnarray}\label{vn}
\mathrm{v_{n,n}} &=& \frac{\sum_{\Delta\phi} C_{r}(\Delta\phi, \Delta\eta)\cos(n \Delta\phi)}{\sum_{\Delta\phi}~C_{r}(\Delta\phi, \Delta\eta)},
\end{eqnarray}
and  then used to extract v$^{even}_{1}$ via a simultaneous fit of v$_{1,1}$ as a function of $p_{T}^{\text {b}}$, for several selections of  $p_{T}^{a}$ with Eq.~\ref{eq:3}:
\begin{eqnarray}\label{corrv1}
\mathrm{v_{1,1}}(\pT^{a},\pT^{b})  &=& \mathrm{v^{even}_{1}}(\pT^{a})\mathrm{v^{even}_{1}}(\pT^{b}) - C\pT^{a}\pT^{b}.
\end{eqnarray}

Here, $C \propto 1/(\langle Mult \rangle \langle p_{T}^{2}\rangle)$ takes into account the non-flow correlations induced by a global momentum conservation~\cite{ATLAS:2012at,Retinskaya:2012ky} and $\langle Mult \rangle$ is the corrected mean multiplicity. For  a given centrality selection, the left hand side of  Eq.~\ref{corrv1} represents the $N \times N$ matrix which we fit with the right hand side using $N + 1$ parameters; N values of v$^{even}_{1}(\pT)$ and one additional parameter $C$, accounting for the momentum conservation~\cite{Jia:2012gu}.  

\begin{figure*}
\centering{
\includegraphics[width=5.5cm,height=8.cm,angle=-90]{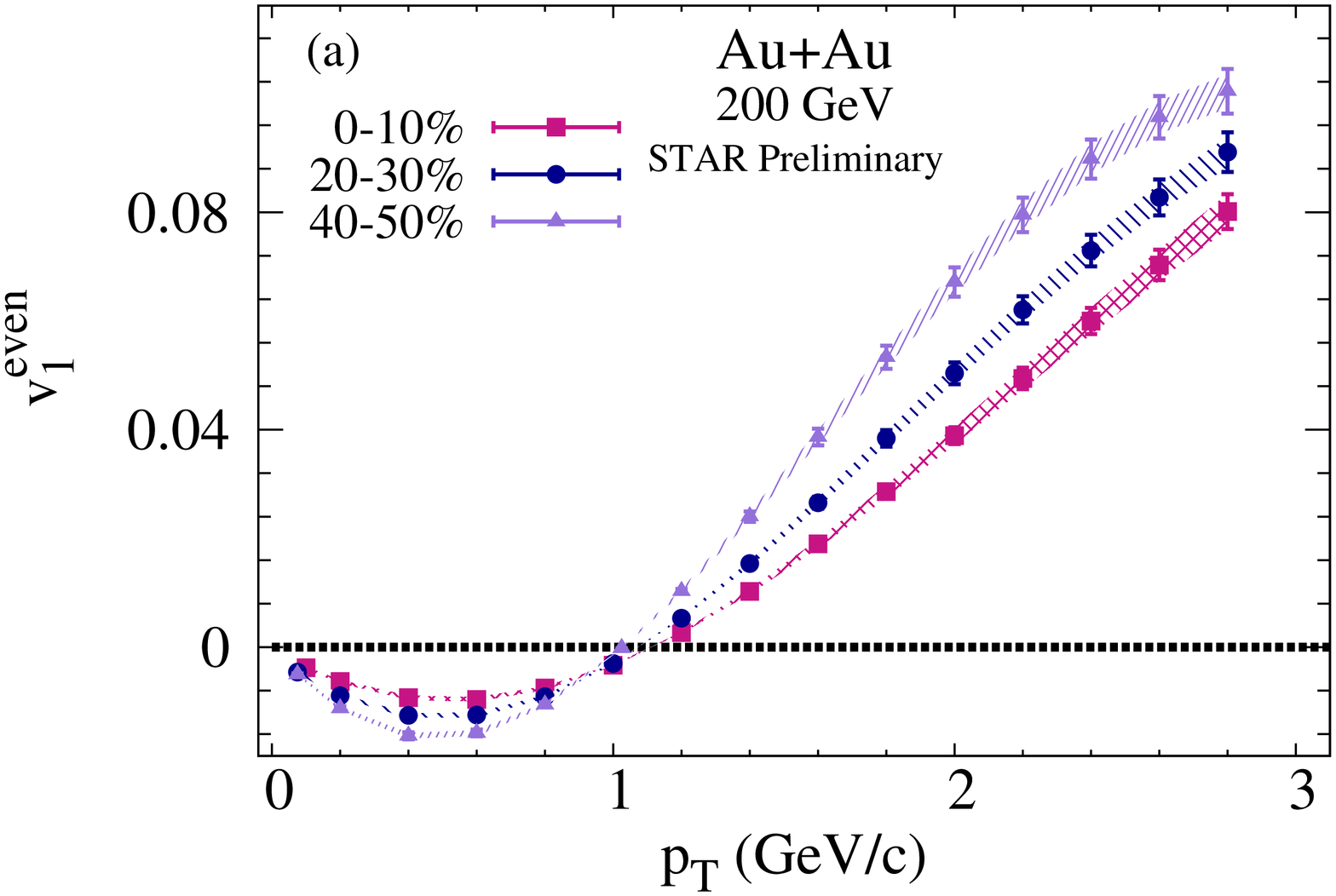}%
\includegraphics[width=5.5cm,height=7.cm,angle=-90]{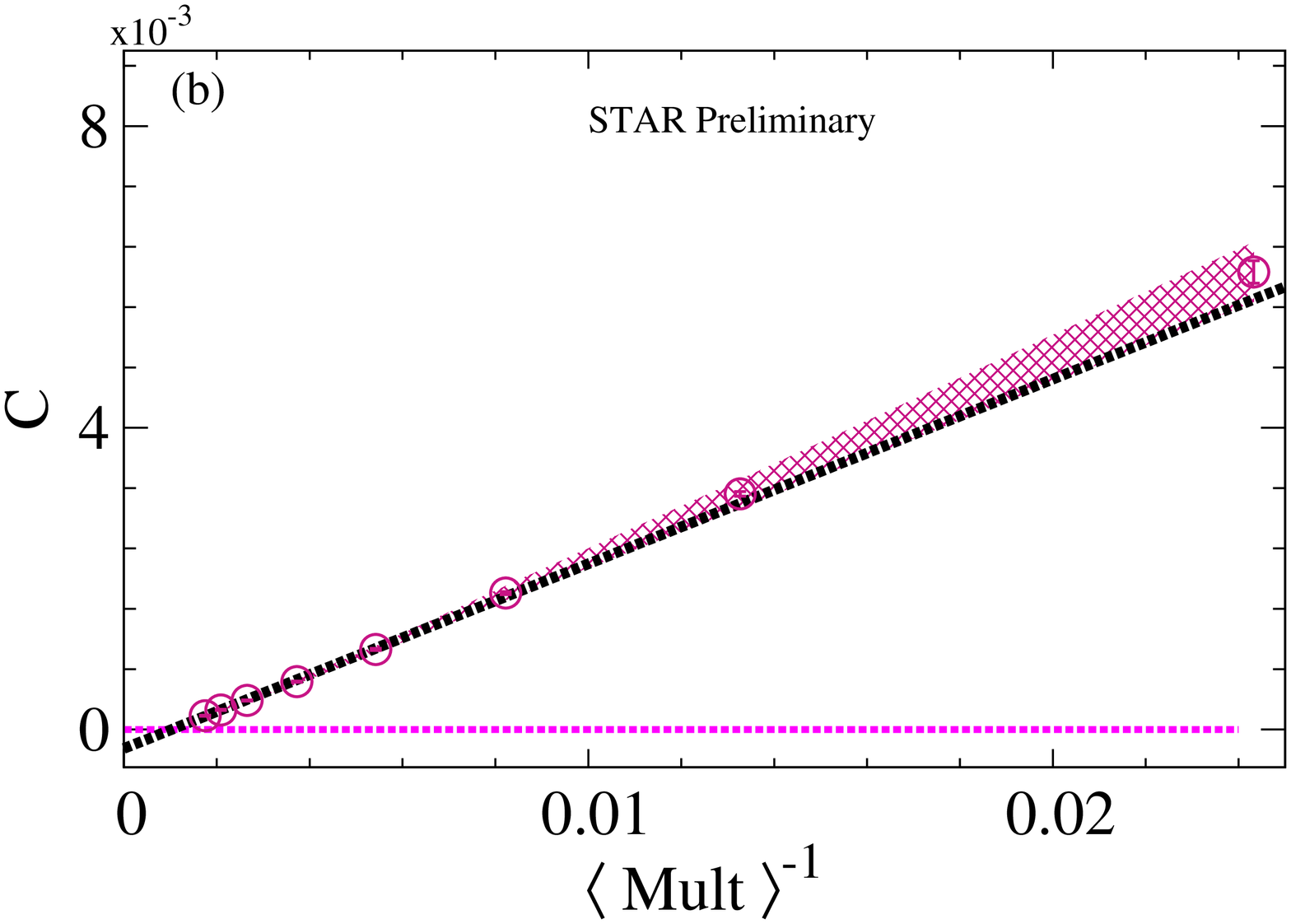}
\caption{(a) The extracted values of v$^{even}_{1}$ vs. $p_T$  for Au+Au collisions at $\sqrt{s_{NN}}~=~ 200$~GeV. (b) A representative set of the associated values of $C$ vs. $\langle Mult\rangle^{-1}$ from the same fits. The shaded bands represent the systematic uncertainty.
 \label{fig1}
 }
}
\end{figure*}

\section{Results}
\label{Results}
\begin{figure*}
\centering{
\includegraphics[width=6.0cm,height=16.cm,angle=-90]{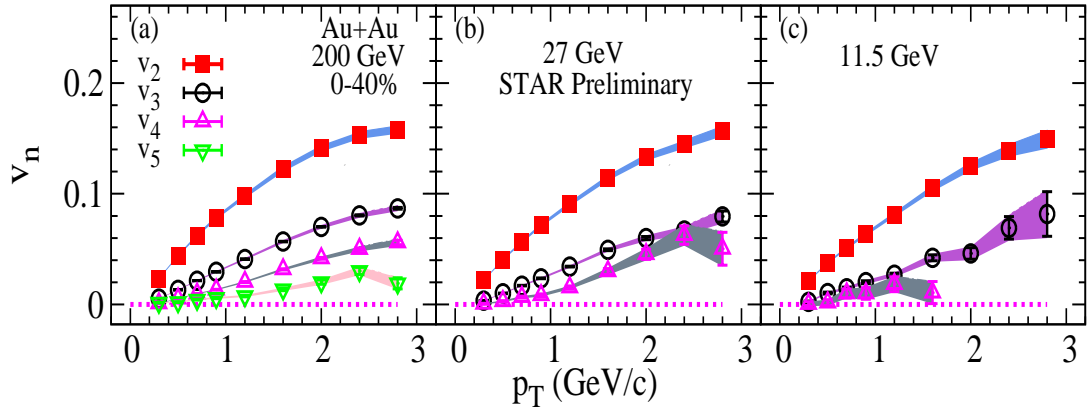}
\caption{Examples of v$_{n}(p_{T})$ as a function of $p_{T}$ for charged particles in 0-40\% central Au+Au collisions. The shaded bands represent the systematic uncertainty.
 \label{fig2}
 }
}
\end{figure*}

\begin{figure*}
\centering{
\includegraphics[width=6.0cm,height=16.cm,angle=-90]{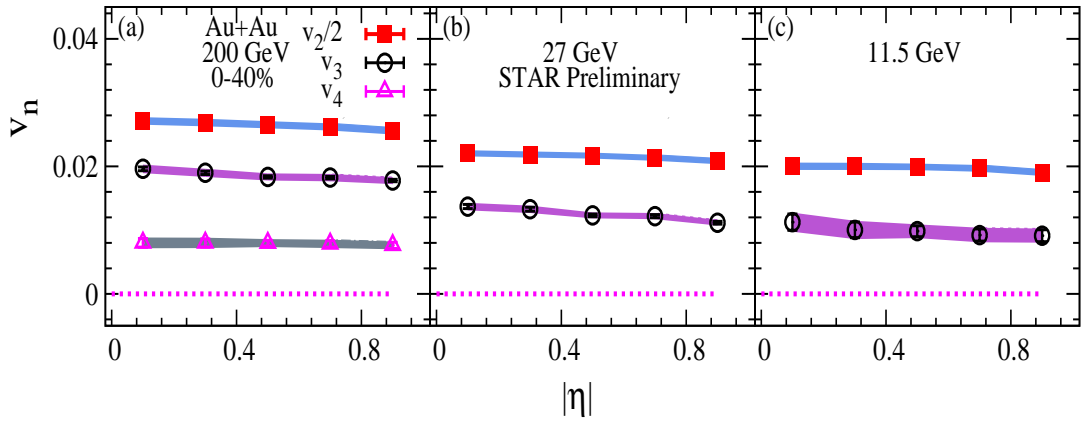}
\caption{Examples of v$_{n}(|\eta|)$ as a function of $|\eta|$ for charged particles in 0-40\% central Au+Au collisions. The shaded bands represent the systematic uncertainty.
 \label{fig3}
 }
}
\end{figure*}

\begin{figure*}
\centering{
\includegraphics[width=6.0cm,height=16.cm,angle=-90]{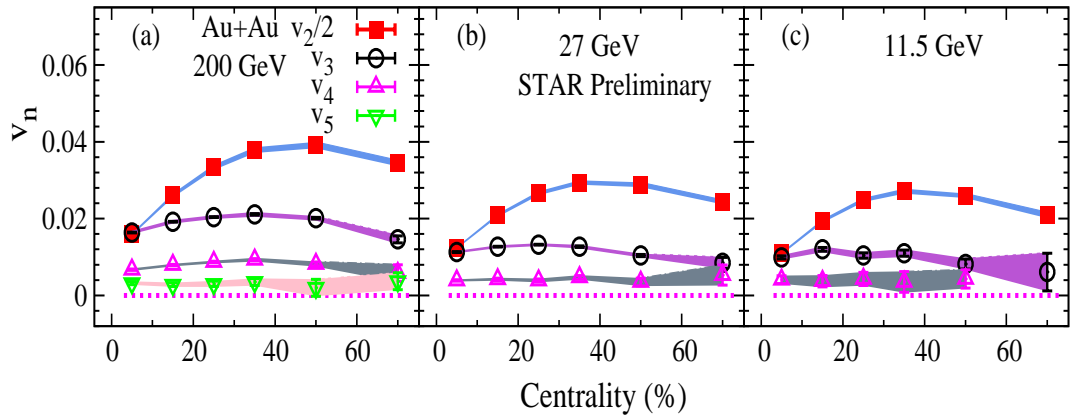}
\caption{Examples of  v$_{n}$(Centrality\%) as a function of Au+Au collision centrality for charged particles with $0.2 < p_{T} < 4$~GeV/c. The shaded bands represent the systematic uncertainty.
 \label{fig4}
 }
}
\end{figure*}

\begin{figure*}
\centering{
\includegraphics[width=6.0cm,height=9.cm,angle=-90]{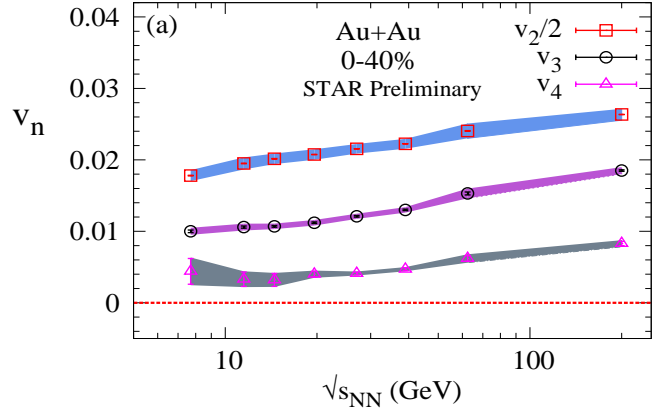}
\caption{Examples of v$_{n}(\sqrt{s_{\rm NN}})$ for charged particles with $0.2 < p_{T} < 4$~GeV/c and 0-40\% central Au+Au collisions. The shaded bands represent the systematic uncertainty.
 \label{fig5}
 }
}
\end{figure*}

Representative results for v$^{even}_{1}$ and v$_{n \ge 2}$ for Au+Au collisions at several different collision energies are summarized in Figs.~\ref{fig1}, \ref{fig2}, \ref{fig3}, \ref{fig4} and \ref{fig5}.
%

The values of v$^{even}_{1}(\pT)$ extracted  for different centrality selections (0-10\%, 20-30\% and 40-50\%) 
are shown  in Fig.~\ref{fig1}(a). They indicate the characteristic pattern of a change from negative  v$^{even}_{1}(\pT)$  at low $\pT$ to positive v$^{even}_{1}(\pT)$ for $p_{T} > 1$~GeV/c. They also show the expected increase of v$^{even}_{1}$ as collisions become more peripheral, in line with the expected centrality dependence of the dipole asymmetry $\varepsilon_1$, where $\varepsilon_1 \equiv  \lvert  (r^3e^{i\phi}) \rvert/r^3$~\cite{Teaney:2010vd,Bozek:2012hy}.  Fig.~\ref{fig1}(b) shows the results for the associated momentum conservation coefficients, $C$; they indicate the expected linear dependence on $\langle Mult\rangle^{-1}$.

Figure \ref{fig2} and \ref{fig3} show $p_T$ and $\eta$ differential $\mathrm{v_{n \ge 2}}$ measurements for the centrality selection 0-40\%, for a representative set of beam energies. Fig.~\ref{fig2} indicates a sizable dependence of the magnitude of v$_{n}$ on $p_{T}$ and the harmonic number, $n$, with similar trends for each beam energy.  Figure~\ref{fig3} shows a similarly strong $n$ dependence for $\mathrm{v_{n \ge 2}}$ but with a much weaker $\eta$ dependence. 

The centrality dependence of v$_{n \ge 2}$ is shown in Fig.~\ref{fig4} for the same representative set of beam energies. They indicate a weak centrality dependence for the higher harmonics, which all decrease with decreasing values of $\sqrt{s_{\rm NN}}$. These patterns may be related to the detailed dependence of the viscous effects in the created medium, which serve to attenuate the magnitude of $v_n$. 
%

Figure~\ref{fig5} shows the excitation functions for the $p_T$-integrated $v_{2, 3, 4}$ for $0-40\%$ central Au+Au collisions. They indicate an essentially monotonic trend for v$_{2}$, v$_{3}$ and v$_{4}$ with $\sqrt{s_{\rm NN}}$, as might be expected for a temperature increase with $\sqrt{s_{\rm NN}}$.

\section{Conclusion} 
In summary, we have performed a comprehensive set of STAR anisotropic flow measurements for Au+Au collisions at $\sqrt{s_{\rm NN}}$ $=$ 7.7-200~GeV. The measurements use the two-particle correlation method to extract the Fourier coefficients, v$_{n > 1}$, and the rapidity-even dipolar flow coefficient, v$^{even}_{1}$. The rapidity-even dipolar flow measurements indicate the characteristic patterns of an evolution from negative v$^{even}_{1}(\pT)$  for $\pT ~<~1$~GeV/c to positive v$^{even}_{1}(\pT)$ for $\pT ~>~ 1$~GeV/c, expected when initial-state geometric fluctuations act along with the hydrodynamic-like expansion to generate rapidity-even dipolar flow. The v$_{n > 1}$ measurements indicate a rich set of dependences on harmonic number $n$, $p_T$, $|\eta|$ and centrality for versus the beam energy. These new measurements may provide additional constraints to test different initial-state models, and to aid precision extraction of the temperature dependence of the specific shear viscosity.

\section*{Acknowledgments}
This research is supported by the US Department of Energy under contract DE-FG02-87ER40331.

\bibliographystyle{JHEP}
\bibliography{BES_v1}

\end{document}